\documentclass[12pt]{article}
\input epsf.sty
\def\be{\begin{equation}} \def\ee{\end{equation}}
\def\bea{\begin{eqnarray}} \def\eea{\end{eqnarray}} \def\ba{\begin{array}}
\def\ea{\end{array}} \def\ben{\begin{enumerate}} \def\een{\end{enumerate}}
  
\def\lll{\label}
\newcommand{\eqn}[1]{(\ref{#1})}

\newcommand{\plb}[3]{Phys. Lett. {\bf B#1} ({#2}) {#3}}
\newcommand{\prd}[3]{Phys. Rev. {\bf D#1} ({#2}) {#3}}
\newcommand{\hepth}[1]{{\tt hep-th/{#1}}}
\newcommand{\hepph}[1]{{\tt hep-ph/{#1}}}

\def\ov{\over}
\def\br{\nonumber\\}

\begin{document}
{}~
\hfill\vbox{\hbox{hep-th/0608032} \hbox{SINP/TNP/06-20}}\break

\vskip .5cm
\centerline{\large \bf
The  $N$-Tachyon Assisted Inflation}
\centerline{\large \bf}

\vskip .5cm

\vspace*{.5cm}

\centerline{ \sc Harvendra Singh}

\vspace*{.5cm}

\centerline{ \it Theory Division, Saha Institute of Nuclear Physics} 
\centerline{ \it  1/AF Bidhannagar, Kolkata 700064, India}

\vskip.5cm
\centerline{E-mail: h.singh[AT]saha.ac.in}

\vskip1cm
\centerline{\bf Abstract} \bigskip

In continuation of the papers hep-th/0505012 and hep-th/0508101 we 
investigate the consequences when  $N$ open-string tachyons  roll 
down simultaneously. We demonstrate that the $N$-Tachyon
system  coupled to  gravity  does indeed
give rise to the assisted slow-roll inflation.

\vfill \eject

\baselineskip=16.2pt

\section{Introduction}
The concept of inflation \cite{guth}, the universe expanding with 
an accelerated expansion, is one of 
the most economic  idea in the early universe 
cosmology. Having an initial inflationary phase solves many puzzles; 
the isotropy and homogeneity of the large scale structure, 
observed flatness 
of the universe and the primordial monopole problem etc. 
Inflation 
is also viewed as the most plausible source of the density 
perturbations in the  universe. 
The conventional inflation models 
deal with a scalar field rolling down in a slowly varying potential in  
FRW spacetime \cite{linde}. Lately now,   
it is also being realised in string theory that the string landscape of 
cosmological 
vacua \cite{kklt} does indeed possess  many inflationary  solutions. In 
fact there is a problem of being
plenty. The string models which directly address the problem of  
inflation are the brane-inflation \cite{dvali} models in which 
brane-antibrane 
force drives the modulus separating them. But the inflation is not 
the slow-roll one. The brane-antibrane inflation 
in warped compactification scenario \cite{kklmmt}, however, lacks precise 
knowledge of the inflaton potential in presence of 
fluxes. The recent idea of $N$ axion inflation 
can be found in 
\cite{kachru,mac}. 
While there are open-string 
tachyon models which can also provide inflation, see 
\cite{anup,sen3,hs1,sinha,hs2}, in general, 
these tachyon models are also  plagued 
with the same large $\eta$ problem as  the conventional models and are 
not favoured for slow-roll inflation, see \cite{kofman}. 
There are number of various 
other attempts to provide stringy inflation models. These are for 
instance inflationary models based on branes intersecting at special 
angles \cite{garcia}, and also the D3/D7 models with fluxes where the 
distance 
modulus plays 
the role of inflaton field \cite{keshav,chen}. The race-track inflation 
driven 
by closed string 
modulus could be found in \cite{racetreck}.

A naive idea of assisted inflation was proposed in 
\cite{liddle} 
in which $N$ scalar fields with identical potentials are considered.
In doing so the Hubble parameter becomes of ${\cal O}(\sqrt{N})$ thus 
increasing the gravitational frictional effects which effectively slow 
down the rolling of the scalar fields.\footnote{ The heterotic 
M-theory assisted inflation models are studied in \cite{becker}.} 
In recent works \cite{hs1,hs2} we studied  tachyon inflation models 
based on $N$ unstable 
D3-branes coupled to gravity. These $N$-tachyon models are found to 
be capable of providing 
solutions with slow-roll inflation provided $N$ is taken sufficiently 
large. In 
fact this requires a critical density of the branes on the 
compact Calabi-Yau 3-fold \cite{hs2}. In this paper we refresh that work 
with a new 
proposal that our model \cite{hs2} with $N$ non-BPS branes is actually 
an 
example of assisted 
inflation involving $N$ tachyons. Again the number density 
has to be large enough for these models 
to be cosmologically viable.  The paper is organised as follows. We review 
the aspects of 
assisted inflation in the section-II. In section-III we study the $N$ 
tachyon assisted inflation. The conclusions are in section-IV.

\section{Review: Assisted Inflation}

The assisted inflation idea was proposed in \cite{liddle} to overcome the 
large 
$\eta$ problem in scalar field driven inflation models. We review the main 
aspects of that work here. We consider a scalar field with   
potential $V(\phi)$  minimally coupled to Einstein gravity \cite{linde}  
\be
\int d^4x \sqrt{-g}[{M_p^2\over 2} R -{1\over 2}(\partial\phi)^2 
-V(\phi)]
\ee
where four-dimensional Planck mass $M_p$ is related to the Newton's 
constant $G$ as
$M_p^{-2}={8\pi G}$. Considering purely time-dependent  field 
in a spatially flat FRW spacetime
\be
ds^2=-dt^2 +a(t)^2\left({dr^2}+r^2(d\theta^2+sin^2\theta
d\phi^2)\right)
\ee
the classical field equations can be written as
\bea\label{dgt1}
&&\ddot\phi=-V,_{\phi}-3H\dot\phi \br
&&H^2={8\pi G\over 3}({\dot\phi^2\ov2}+V)
\eea
where $H(t)\equiv\dot a/a$ is the Hubble parameter. This simple model has 
proved to be a prototype for explaining the mechanism of  
inflation in early universe. For example, if we take a quadratic 
potential $V=m^2\phi^2/2$ and let the field roll down from some large 
initial value, the field will roll down to its minimum value $\phi=0$ and 
 so spacetime will inflate \cite{linde}. But the 
inflation has to be a slowly 
rolling one in order to fit with cosmological observations. That is 
the field $\phi$ must vary slowly such that 
there is a vanishing acceleration,  $ \ddot\phi\sim0$. Under the slow-roll 
conditions,
the time variation of $\phi$  gets 
related to the slop of the potential as
$$\dot\phi\simeq-{V,_{\phi}\over 
3 H}.$$
So the potentials with gentle slop are preferred for a good 
inflation. 

The standard slow-roll parameters are \cite{sasaki}
\be
\epsilon\equiv{ M_p^2\over2} {V'^2\over V^2},~~
\eta\equiv{ M_p^2} {V''\over V}
\ee
where primes are the derivatives  with respect to $\phi$. For the 
slow-roll inflation 
both $\epsilon$ and $\eta$ will have to be small. Also these parameters 
are in 
turn related to the spectral index, $n_s$, of the scalar density fluctuations
in the early universe as 
\be n_s-1\simeq -6\epsilon+2\eta.
\ee 
A nearly uniform power spectrum observed over a wide range of 
frequencies in the density 
perturbations in CMBR measurements \cite{wmap}, however, requires  
$n_s\simeq .95$. It can 
be achieved only if $$\epsilon\ll1,~~~\eta\ll 1.$$
These are some of the stringent bounds from cosmology which inflationary 
models have to 
comply with. 

For single scalar field the power spectrum of the scalar curvature 
perturbations can be written as \cite{sasaki}
\be
P_R={1\over 12\pi^2M_p^6}{V^3 \over V'^2} 
\ee
The  amplitude (or size) of these fluctuations are governed by 
\be\label{df1}
\delta_H={2\over5}\sqrt{P_R}={1\over 5\sqrt{3}\pi M_p^3}{V^{3/2} \over 
V'} \le 2\times 10^{-5}
\ee
The inequality on the right side of equation \eqn{df1} indicates the 
COBE 
bound on the size of these 
perturbations at the beginning of the last 50 e-folds of 
inflation.\footnote{ The number of e-folds, $N_e$, during inflationary 
time interval $(t_f-t_i)$ are estimated as $N_e=\int_{t_i}^{t_f} H dt$. 
Our 
universe requires 50-60 e-folds of expansion in order to explain the 
present size of the observed large scale structure.}  

It can be easily seen that
 the models with quadratic potentials are not useful for 
inflation as they are plagued with so called $\eta$-problem. From the 
above we find that $\eta\sim 2 M_p^2/\langle\phi\rangle^2$. Therefore
 $\eta$ will be small only if $\phi$ has trans-Planckian vacuum 
expectation value at the time when inflation starts. But 
allowing the quantum fields to have  
trans-Planckian vevs will spoil the classical analysis and will involve 
quantum corrections. However, one can also see recent developments in
\cite{mac}. 

An effective resolution of the large $\eta$ problem can come 
from assisted inflation model \cite{liddle}. The model  involves 
large number of scalar 
fields $\phi_i$ $(i=1,2,\cdots,N)$. Let us demonstrate it here for a 
simple quadratic 
potential $m^2\phi^2/2$. We take the case where all scalars have the 
same mass $m$ ($i.e.$  potential). The scalar fields are taken to be 
noninteracting 
but  are minimally coupled to 
gravity. The equations of motion are
\bea\lll{aeqn1}
&&\ddot \phi_i= -{m^2 \phi_i}-3 H \dot \phi_i\ ,\br
&& H^2= {8\pi G \over 3 }\sum_{i=1}^N (V(\phi)+ {\dot 
\phi_i^2\over2})\ .
\eea
A particular solution exists where all fields are taken equal
$\phi_1=\phi_2=\cdots=\phi_N=\Phi$. The simplified equations  become
\bea\lll{aeqn2}
&&\ddot \Phi= -{m^2 \Phi}-3 H \dot \Phi\br
&& H^2= {8\pi G \over 3 } N (V(\Phi)+ {\dot 
\Phi^2\over2})
\eea
One finds that eqs.\eqn{aeqn2} are having an effective Newston's 
constant  $G.N$ when compared with eqs.\eqn{dgt1}. So 
we easily calculate that $$\epsilon=\eta={2 M_p^2\over N \Phi^2} \ .$$ 
So if $N$ is sufficiently large enough $\Phi$ need not have 
trans-Planckian vev.
It can also be seen that since $H\sim {\cal O}(\sqrt{N})$ it can provide 
slow-roll 
inflation since \cite{liddle}
 \be
n_s-1=2{\dot H\over H^2}\sim {\cal O}({1\over N})\ .
\ee
If larger is the value of $N$, more flat will be the observed spectrum. 
As an estimate, for 
$$ {2 M_p^2\over N \Phi^2}\sim 
.01$$
and the scalar mass $$m\sim 10^{-3} M_p\, ,$$ one can
can produce good slow-roll inflation.  In fact, in Ref.\cite{liddle}
it was shown that the same phenomenon occurs  even for exponentially fast 
potentials like $V=V_0\, exp(-\alpha\phi_i)$.

 \section{$N$-Tachyon Inflation}
In this section we shall 
demonstrate that the assisted inflation can also occur for open string 
tachyons. It is rather 
natural for tachyons, unlike in the scalar field model above,
to have the same masses $-{M_s^2/2}$ and also the identical potentials. 

We consider  $N$ unstable D$3$-branes distributed over a 
compact 
$CY_3$ with the six-volume parametrised as 
\be\label{vol}
{\cal V}_{(6)}= (l_s)^6 v_0 
\ee
where 
$l_s=\sqrt{\alpha'}=1/M_s$ is the 
string length. In the dilute `brane gas' approximation (far separated 
branes in a large six-volume ($v_0\gg1$))
 the unstable branes will have leading tachyon modes coming from stretched 
open strings attached to  themselves only. 
Thus we are ignoring the tachyonic modes which
come from the strings attached between a pair of far separated 
branes. Effectively
we are assuming that the branes are  non-interacting but 
interact through the gravity only. So the 
system is reduced to 
exactly $N$ tachyon fields with gravitational interaction. The 
effective low energy tachyon 
action can
be written from
 \cite{sen0,sen1}
\be\lll{tachyact}
-\sum\limits_{i=1}^N\int d^4x\, V_i(T_i)\sqrt{-{\rm det} (g_{\mu\nu}+ 
\alpha'\partial_\mu
T_i\partial_\nu T_i)} \ee   
where  $T_i$'s are the $N$ tachyon
fields, $V_i(T_i)$ is the tachyon potential for $T_i$, and
$g_{\mu\nu}$ is the pull-back of the spacetime metric. There is no 
summation over index $i$ within the integrand.

From the Sen's conjectures the
tachyon potential, $V(T)$, is symmetric and
has a central maximum at $T=0$ with
a pair of global minima at $T\to\pm\infty$. 
The exact form of
tachyon potential is not quite well known, although it is
known to behave as $$V(T)\sim e^{-{T\over \sqrt{2}}}$$ near the tachyon
vacuum ({\it  non-perturbative open string vacuum}) at $T=\infty$. While 
near the 
perturbative  vacuum at $T=0$, the region of our interest, 
the potential 
can  precisely be 
written as
\be\label{pot1}
V(T)=\sqrt{2}\, \tau_3(1-{T^2\over 4})
\ee
where $\tau_3={1\over (2\pi)^3 g_s \alpha'^2}$ is the BPS D3-brane 
tension, 
$g_s$ being the string coupling constant.
The potential \eqn{pot1} is valid for $|T|\ll 2$ only.
It is like an inverted oscillator and the small excitations  
near $T=0$ are tachyonic with negative $(mass)^2=-{M_s^2\over2} 
$. 

In the flat spacetime background, the equations
of motion 
for the purely time-dependent (homogeneous) tachyon fields can 
be written from \eqn{tachyact}
\bea\lll{ta1}
&&\ddot T_1=-(1-\alpha' \dot T_1^2) {M_s^2\over  V_1} {d V_1\over 
dT_1} \ ,
\br &&\ddot T_2=-(1-\alpha' \dot T_2^2) {M_s^2\over  V_2} {d V_2\over 
dT_2}\ ,
\br && \vdots
\eea
These equations  are completely decoupled which is our primary assumption. 
So
we can  find  a rolling tachyon solution where all  $N$ tachyon fields 
roll down simultaneously. We can have 
$T_1(t)=T_2(t)=\cdots=T_N(t)=\phi(t)$
provided 
we also take $V_1(T_1)=V_2(T_2)=\cdots=V_N(T_N)=V(\phi)$. This is  
guaranteed from the 
known form of tachyon potential \eqn{pot1} valid in the perturbative
open string  
vacuum. Under this {\it simultaneous} ansatz, the effective 
action \eqn{tachyact} will simply become
 \be\lll{tachyact1}
-N\int d^4x  V(\phi) \sqrt{-g} \sqrt{1 + \alpha '
(\partial_\mu \phi)^2} \ . 
\ee   
Note that this was the action which we called as the {\it fat} 
unstable 3-brane in 
\cite{hs1,hs2}.
We shall now couple this system to four-dimensional Einstein gravity
\be\lll{dgf}
{M_p^2\over 2}\int d^4x \sqrt{-g} R \ .
\ee
The Planck mass $M_p$ is related to ten-dimensional string mass as
\be\lll{mpms} 
M_p^2={ v\over g_s^2}M_s^2\ ,
\ee
where our convention has been 
\be \kappa_{(10)}^2=(2\pi)^7g_s^2\alpha'^4\ , ~~~ 
 v\equiv v_0/(2\pi)^7\ .
\label{con1}\ee  
For the effective 
four-dimensional 
 analysis to remain valid, we must take string coupling $g_s\ll 1$ and 
also $v_0\gg1$ in order 
to suppress the stringy corrections. From the relation \eqn{mpms},
this always brings the string mass $M_s$  much lower than $M_p$, 
which is a common assumption in large volume ($v_0\gg 1$) string 
compactifications. 

 In order to study cosmological solutions we take FRW ansatz and consider 
purely time dependent tachyon fields.
From \eqn{tachyact} and \eqn{dgf} the combined gravity and tachyon field 
equations, keeping the simultaneous ansatz as above,  
can be written as 
\bea\lll{teqn}
&&\ddot \phi=-(1- \alpha'\dot \phi^2)\left(M_s^2 {V,_{\phi}\over 
V}+3 H \dot 
\phi \right) \br
&& H^2= {8\pi G\over 3 }  
(1-{\phi^2\over 
4}){\sqrt{2} \tau_3 N\over\sqrt{1- \alpha' \dot \phi^2}} 
\eea
We again emphasize that these equations are valid only near the top of 
the 
tachyon potential where $\phi^2\ll 4$ and 
$\dot\phi^2\ll M_s^2$, these follow from the tachyon potential 
\eqn{pot1} which is 
valid only near $T\sim 0$. Using these approximations and keeping upto 
quadratic terms in the field $\phi$, the equations 
\eqn{teqn} can be approximated as
\bea\lll{teqn1}
&&\ddot \phi= ({M_s^2 \over 
2}\phi-3 H \dot 
\phi ) + {\cal O}(\phi^3) \br
&& H^2= {8\sqrt{2}\pi G\tau_3N\over 3 }(1-{\phi^2\over4}+ \alpha'{\dot 
\phi^2\over2}) +{\cal O}(\phi^4) \ .
\eea
Note that the field $\phi$ is dimensionless so far in our analysis.
We can restore its  canonical mass dimension by the rescaling 
$\phi\to\sqrt{\alpha'}\,\phi$. In which case, dropping the higher 
order terms altogether, we get
\bea\lll{teqn2}
&&\ddot \phi= {M_s^2 \over 
2}\phi-3 H \dot \phi  \br
&& H^2= {8\pi G \over 3 }{\bar N \over g_s}(V_{eff}+ {\dot 
\phi^2\over2})
\eea
where we defined
\be
V_{eff}=M_s^4-M_s^2{\phi^2\over4},~~ \bar 
N\equiv{\sqrt{2} N\over(2\pi)^3 } \ .
\ee
These are the actual equations relevant for an assisted inflation, find 
the similarity with eqs. 
\eqn{aeqn2}. 
Comparing eqs. \eqn{aeqn2} and \eqn{teqn2} we determine that the mass 
of the inflaton field 
is the tachyon mass $-M_s^2/2$. But unlike conventional quadratic 
potential $m^2 \phi^2/2$,  
the potential $V_{eff}$ has a maximum at $\phi=0$. In any case, for 
very large value of $N$, we will have 
$H={\cal O}(\sqrt{N})$ and the slow-roll 
assisted inflation is possible. This is our main result.

\section{Slow-roll parameters}

In order to know the slow-roll parameters, it will be useful to define a new 
field,
\be\label{sl1}
\psi\equiv ({\bar N\over g_s})^{1\over2}\phi
\ee
In terms of $\psi$ the equations in \eqn{teqn2} become
\bea\lll{teqn3a}
&&\ddot \psi= {M_s^2 \over 2}\psi-3 H \dot\psi  \br
&& H^2= {8\pi G \over 3 } (V+ {\dot 
\psi^2\over2})
\eea
where 
\be\label{veff}
V={\bar N\over g_s}M_s^4-M_s^2{\psi^2\over4}.
\ee
We must caution the reader here, though it appears that there 
is only one scalar field in eqs. 
\eqn{teqn3a},
one must keep in mind that actually there are $N$ of them. The energy 
density for single unstable brane is $\sqrt{2} \tau_3$. 
Viewing 
\eqn{teqn3a} as a single scalar field system the
inflationary slow-roll
parameters remain 
$$\epsilon\equiv{ M_p^2\over2} {V'^2\over V^2},~
\eta\equiv{ M_p^2} {V''\over V}$$
with derivatives with respect to the new field $\psi$.
These become
\bea
&&\epsilon= {M_p^2\over2}({-M_s^2\psi\over 2V})^2\simeq{1\over2}{v\over 
\bar N g_s}({\phi\over 2M_s})^2 \br
&&\eta\simeq -{1\over2}{v\over \bar N g_s}
\eea
where we have used eq.\eqn{sl1} and the fact that  $\phi \ll 2 M_s$. Thus, 
already 
$\epsilon \ll 1$ 
provided ${\bar N \over v}={\cal O}(1/g_s)$. This is a quality reminiscent 
of topological inflation where inflation happen very close to the top of a 
potential. 
But in order to make $\eta\ll 1$, we need to have 
\be\label{dfo}
{\bar N\over v}\gg {1\over g_s}\ .
\ee 
Note that unlike trans-Planckian 
vevs in scalar field inflation models, in the case of tachyon field 
the inflation happens when
$\langle\phi\rangle \ll 2 M_s\ll M_p$.

We now determine the  size of scalar density fluctuations 
\be\lll{delh}
\delta_H={1\over \sqrt{75} \pi}{1\over M_p^3} {V^{3\over2}\over 
V,_{\psi}}\ .
\ee
Evaluating this quantity using eqs. \eqn{veff} and 
\eqn{sl1}, we find
\be\label{delh1}
\delta_H={\sqrt{2}\over \sqrt{75} \pi}{M_s\over M_p} {g_s N\over 
(2\pi)^3 
v}{2M_s\over \phi}
\ee
Thus, by taking 
\be\label{data1}
{M_s\over M_p}={g_s\over\sqrt{v}}\approx 10^{-6},~ {\sqrt{2}g_s 
N\over (2\pi)^3 v}\approx 10
\ee
we estimate from \eqn{delh1}
\be
\delta_H\sim 3.7\times 10^{-7}  ({2M_s\over \phi})
\ee
That is, the size of amplitudes will be  $\sim 1.8 \times 10^{-5}$ if we 
choose 
${\phi\over 2M_s}=.02$. In summary,
we can get the size of density fluctuations within the COBE bound 
$\delta_H\le 1.9\times 
10^{-5}$ at the start of the last 50 
e-folds of inflation. Note that the inflation will end before 
$\langle\phi\rangle \sim 2 
M_s$. 
\vskip.5cm
\noindent\underline{\bf Critical number density}: 

From equations \eqn{dfo} and \eqn{data1} it is clear that we need 
to have at least
\be 
{\sqrt{2}N \over( 2\pi)^3 v}\ge {10\over g_s}\ .
\ee 
Using the definition $v\equiv v_0/(2\pi)^7$ we get 
\be\lll{data2}
{N\over v_0}\ge {5\sqrt{2}\over (2\pi)^4 g_s}\ .
\ee
This is what we call the 
critical  number density of the unstable D3-branes on the compact manifold 
for which $v_0\gg 1$. Anything less than that will not work.
Note that $v_0$ is the absolute measure of the $CY_3$ volume in $(l_s)^6$ 
units.
So if we take reasonably smaller value of the string coupling, $g_s\sim 
.1$, we get an estimate  
\be
{N\over v_0}\ge .045\ .
\label{data3}
\ee
The eqs. \eqn{data2} and \eqn{data3}  do seem to  define for us 
a very dilute gas of D3-branes on the $CY_3$. 
Eq.\eqn{data3} shows that we need to have less than
one D3-brane per unit 
six-volume of the compact manifold measured in string length units.  
This justifies our earlier assumption
that inter-brane interactions could 
be ignored and the $N$ tachyon action eq.\eqn{tachyact} can 
be taken as a
direct sum of the action for individual tachyons.  

\vskip.5cm
\noindent\underline{\bf Some Remarks}: 

In the above we took $g_s\sim .1$ while any value in the range $.01\le 
g_s\le .1$ is allowed. But if $g_s$ is taken 
further smaller than $.01$, the ratio
$N/v_0$ will get far away from a dilute gas approximation since we 
would like to keep not more than one 3-brane per unit 
Calabi-Yau six-volume. 
It perhaps suggests us that we should then include the 
warping of the spacetime in our analysis something in line 
with the works \cite{kklmmt,sinha}. 
It was shown 
in Ref.\cite{sinha} that the warping effects can produce slow-roll 
inflation.     
There are important issues of moduli stabilisation 
during inflation, the moduli fields could intefere 
with an otherwise successful inflation. 
We have ignored these effects in this 
paper as those should be addressed separately.

Let us also estimate the value of the Hubble parameter at the beginning 
of the inflation ($\psi\sim 0$). Using eqs. \eqn{teqn3a} and \eqn{data1}
\be
H(0)^2\simeq {8\pi G\over 3}{\bar N\over g_s} M_s^4 \approx {10\over 3} 
M_s^2\ .
\ee
It means that $H(0)\sim M_s$, 
 which is good because for a low energy analysis to remain 
valid $H\le M_s$. 
Also from \eqn{data1}, we get the string scale of $10^{13} GeV$ much 
smaller than $M_p$. 
This is an usual feature of large volume compactifications. For 
$g_s\sim .01$, we estimate the volume parameter $v\sim 10^8$. The 
absolute $CY_3$ volume ${\cal V}_{(6)}$
measured in string length units will be $(2\pi)^7 10^8 (l_s)^6$. Equating 
${\cal V}_{(6)}\equiv (2\pi R)^6$, it
gives the compactification radius $R \sim 29 l_s$, which makes the 
compact volume reasonably large to suppress higher order string effects.

 \section{Conclusion}
We have studied in detail the $N$-tachyon inflation in an FRW spacetime. 
It is shown
that when a large number of tachyon fields roll down 
simultaneously from the top of the tachyon potential, we get an 
assisted slow-roll 
inflation. This is also the concrete example of assisted inflation in 
string theory involving only tachyons. 
The  results of this work match with 
our previous 
work \cite{hs2}. Particularly, we need to have a dilute number density of 
unstable D3-branes
on the compactified manifold such that there are less than one D3-brane 
per unit string size volume of the compact manifold. We have not 
considered the
effects of warping in the throat regions of the compact manifold. If these 
effects are considered
together with sizable number of unstable D3-branes, we will be assured 
of slow-roll inflation with desired properties so as to fit the 
cosmological bounds.

\leftline{\it Acknowledgments:}

I wish to thank M. Haack, L. McAllister for helpful discussions 
and specially thank M. 
Zagermann for useful discussions and comments on the draft. 
The assistance and hospitality under the  Associateship 
Programme of ICTP has been crucial during the course of this work. 
This work got partly shaped up during the 
visits to  Max-Planck-Institute, Munich and 
II Institute, University-of-Hamburg for which
I am grateful to M. Zagermann and J. Louis for warm 
hospitality. 

\vskip.5cm
\noindent {\it Note added}: After we reported this work we came to know 
of Ref.\cite{cai} where the authors study 
assisted inflation with exponential tachyon 
potentials, which is valid for large values of tachyon field $T$. 
While we have considered the potential with a central 
maximum which is valid near small $T$.

\end{document}